\documentclass[12pt]{article}
\usepackage{latexsym}

\usepackage{amssymb}
\usepackage{amsmath}

 \usepackage{epsfig}
 \usepackage{graphicx}
 
\usepackage{cite}
\usepackage{hyperref}

\usepackage{tikz}
\usepackage{tikz-3dplot}   

\numberwithin{equation}{section}  

\newcommand{\be}{\begin{equation}}
\newcommand{\ee}{\end{equation}}
\newcommand{\ba}{\begin{eqnarray}}
\newcommand{\ea}{\end{eqnarray}}
\newcommand{\baa}{\begin{array}}
\newcommand{\eaa}{\end{array}}
\newcommand{\bi}{\begin{itemize}}
\newcommand{\ei}{\end{itemize}}
\newcommand{\edoc}{\end{document}}

\newcommand{\nn}{\nonumber \\}
\newcommand{\nr}[1]{(\ref{#1})}
\newcommand{\la}[1]{\label{#1}}

\newcommand{\rmi}[1]{{\mbox{\scriptsize #1}}}
\newcommand{\fr}[2]{{\frac{#1}{#2}\,}}
\newcommand{\fra}[2]{{\textstyle{\frac{#1}{#2}\,}}}  
\newcommand{\mn}{{\mu\nu}}
\newcommand{\bfx}{{\bf x}}
\newcommand{\bfy}{{\bf y}}

\newcommand{\bfv}{{\bf v}}
\newcommand{\bfu}{{\bf u}}

\newcommand{\p}{{\partial}}

\def\Tr{{\rm Tr\,}}

\def\CO{{\cal O}}

\def\CD{{\cal D}}
\def\gsim{\raise0.3ex\hbox{$>$\kern-0.75em\raise-1.1ex\hbox{$\sim$}}}
\def\lsim{\raise0.3ex\hbox{$<$\kern-0.75em\raise-1.1ex\hbox{$\sim$}}}

\begin{document}

\begin{flushright}
HIP-2020-34/TH\\
\end{flushright}

\begin{center}

\centerline{\Large {\bf Memory effect in Yang-Mills theory with an axion}}

\vspace{8mm}

\renewcommand\thefootnote{\mbox{$\fnsymbol{footnote}$}}
Niko Jokela,${}^{1,2}$\footnote{niko.jokela@helsinki.fi}
K. Kajantie,${}^2$\footnote{keijo.kajantie@helsinki.fi} and
Miika Sarkkinen${}^{1}$\footnote{miika.sarkkinen@helsinki.fi}

\vspace{4mm}
${}^1${\small \sl Department of Physics} and ${}^2${\small \sl Helsinki Institute of Physics} \\
{\small \sl P.O.Box 64} \\
{\small \sl FIN-00014 University of Helsinki, Finland} 
 
 \vspace{0.8cm}
\end{center}

 \vspace{0.8cm}
 
\setcounter{footnote}{0}
\renewcommand\thefootnote{\mbox{\arabic{footnote}}}

\begin{abstract}
\noindent 
We study the empirical realization of the memory effect in Yang-Mills theory with an axion-like particle, especially in view of the classical vs. quantum nature of the theory. We solve for the coupled equations of motion iteratively in the axionic contributions and explicitly display the gauge invariant effects in terms of field strengths. We apply our results in the context of heavy ion collisions, in the thin nuclear sheet limit, and point out that a probe particle traversing radiation train acquires a longitudinal null memory kick in addition to the usual transverse kick. 
\end{abstract}

\newpage


\section{Introduction}
Yoshida and Soda have studied \cite{Yoshida:2017fao} how a possible cosmological axion background would affect measurements of the electromagnetic memory effect \cite{winicour,bieriED,stromED,Mao:2017wvx,Hamada:2017atr,Hamada:2018cjj}. Not surprisingly, a new radiation mode became observable. On the other hand, the extension of electromagnetic memory to non-Abelian theories has been studied in \cite{Pate:2017vwa,ymmemo2,Jokela:2019apz,Campoleoni:2019ptc}. The purpose of this article is to study how non-Abelian memory would be affected by a simultaneous excitation of a color singlet axion-like (called axionic in the following) degree of freedom.

Physically, non-Abelian theories develop a gap and do not propagate as massless radiation. Nevertheless, classical radiation-like color field configurations appear in the framework of analysing the dynamics  of collisions of ultrarelativistic large nuclei in terms of the McLerran-Venugopalan model \cite{Ayala:1995kg} and color glass condensate (CGC) \cite{iancu}. A single collision can be interpreted as a burst of classical non-Abelian radiation for which the memory effect can be formulated. However, to obtain physical gauge choice independent results one has to average over an ensemble of color field configurations.

We shall set up the problem by studying the case in which there is one large nucleus, the wave function of which is excited by a weak probe, like a single nucleon. We assume an axionic degree of freedom is also excited, write down the coupled equations of mode and solve \cite{Gelis:2005pt} the fluctuations of the gauge fields induced by the axion and finally compute the effect on the memory, total transverse kick of a test particle. The outcome is that there on the classical level indeed is a new parity violating mode, in analogy with \cite{Yoshida:2017fao}. However, when going over to quantum theory, the effect averages itself out in the infinitely contracted nucleus limit. We list several effects which could contribute in
the finite width limit but are so far unable to compute them. These can be addressed in future studies.

In cosmology, the motivation for studying an axionic background is, for example, dark matter. In non-Abelian field theory, QCD,  the motivation is the anomalous non-conservation of the axial U(1) current. There is an extensive literature on the appearance of these phenomena in nucleus-nucleus collisions \cite{Kharzeev:2015kna}.  Axion-like effects can appear in the single nucleus case in a very subtle way in polarized deep inelastic scattering \cite{Tarasov:2020cwl}. Non-Abelian gauge fields together with axions appear also in studies of inflationary cosmology \cite{Maleknejad:2012fw,Lozanov:2018kpk}.

The rest of this paper is organized as follows. Coupled equations of motion and solution iterative in the axion are written down in Section~\ref{eomsec}; leading order equations are solved in Section~\ref{O0}, and next-to-leading ones in Section~\ref{O1}, at the end of which axion-corrected gauge fields are summarized. 
Effect on the memory is derived in  Section~\ref{memo} and Section~\ref{sec:conclusions} contains our conclusions. Two appendices contain additional details.

\medskip

\paragraph{Conventions}
Color conventions are 
$D_\mu=\p_\mu-igA_\mu\equiv D_\mu(A)$ , $A_\mu=A_\mu^a T_a$, $[T_a,T_b]=if_{abc}T_c$, $a,b,c=1,...,N_c^2-1$,
$F_\mn=i/g [D_\mu,D_\nu]=\p_\mu A_\nu-\p_\nu A_\mu-ig[A_\mu,A_\nu]\equiv F_\mn(A)$. Under
a unitary gauge transformation $U(x)$ 
\be
A_\mu\to A_\mu^\prime=UA_\mu U^\dagger+i/g\, U\p_\mu U^\dagger,\quad
F_\mn\to F_\mn^\prime=UF_\mn U^\dagger.
\la{gt}
\ee 
The adjoint representation is defined by $(T^a)_{bc}=-if_{abc}$. For commutators of color matrices $M=M_a T^a$ in any representation we use $[M,N]_c=if_{abc}M_aN_b\equiv M^\rmi{adj}_{cb}N_b=(M^\rmi{adj}N)_c$, where in the RHS $M^\rmi{adj}$ is in adjoint representation and $N=(N_b)$ is a color vector. In a related projection we may have a matrix equation $UMU^\dagger = J$ in any representation, $M,J$ are Lie algebra elements, $M=M_a T_a$, $\Tr T_aT_b=T_R\delta_{ab}$. Then the $a-$component of the equation is
\be
 J_a = \fra1{T_R}\Tr[T_aUT_bU^\dagger]M_b\equiv C(U)_{ab}M_b \la{cu}
\ee
and the equation is written as a matrix$\times$vector equation $C(U)M=J$ with a new adjoint matrix $C(U)$. The transformation $C(U)$ forms a representation of the group in the sense that $C(UV)=C(U)C(V)$. However, if $U,\,T_a$ are in the adjoint representation, the new matrix is exactly the same as the original $U$ matrix, 
\be
 C(U)=\fra1{T_R}\Tr[T_aUT_bU^\dagger]=U_{ab}\ . \la{uadj}
\ee 
This is easy to verify infinitesimally, by writing $U\approx 1+i\theta_a T_a$ and doing the trace. The matrix equation $UMU^\dagger = J$ then has become a matrix$\times$vector equation $UM=J$.

The metric convention is using mostly plus light cone coordinates, $v\equiv v^\mu =(v^+,v^-,\bfv)$, $v^+ = \fra1{\sqrt2}(v^0+v^1)=-v_-$, $v^i=v_i$, $v\cdot u=-v^+u^--v^- u^++\bfv\cdot\bfu$, and $v_T=|\bfv|$. Here we have taken $x^1=x_1=x_L$ as the longitudinal coordinate, $x^i=x_i=(x^2,x^3)$ are then the transverse ones.

\section{Equations of motion}\la{eomsec}

The action of $N_f=0$ QCD with a pseudoscalar axion $\chi$ is 
\be
 S[A^\mu_a,\chi]=\int d^4x\left[-\fra14 F_\mn^a F^\mn_a-\fra{\lambda}4 \chi F_\mn^a \tilde F^\mn_a-\fra{f^2}2\p_\mu\chi \p^\mu\chi - V(\chi)+J_\mu^a A^\mu_a \right] \ .
\la{sax}
\ee
We focus the attention on the coupling of axions with the gluon sector and omit quarks from consideration.
Here $F^a_\mn$ is the field tensor of SU($N_c$) Yang-Mills theory, $\tilde F^a_\mn$ its dual, $\lambda$ is a dimensionless parameter counting the number of axion interactions, $f$ is a parameter of dimension 1, $V(\chi)$ is axion potential, often chosen as $ \mu^4(1- \cos\chi)$ to retain some shift symmetry and  $J$ is a color current. Without the axion this action and its extensions has been used to study \cite{ymmemo2,Jokela:2019apz} the possibility and even phenomenology of YM memory \cite{Pate:2017vwa,Campoleoni:2019ptc} in heavy ion collisions in which particularly large densities of gluons and thus classical gluon fields are involved. The aim of the present study is to investigate how  possible existence of an axionic interaction would affect these considerations.

In the cosmological context the effect of a cosmic axion background on the usual U(1) electromagnetic memory has been studied in \cite{Yoshida:2017fao,Mao:2017wvx,Hamada:2018cjj}. While the usual memory is of $E$-type \cite{winicour}, parity breaking properties of the axion lead also to the appearance of $B$-type memory. Interactions between YM fields and axions have also been studied in cosmology in the context of inflation \cite{Maleknejad:2012fw,Lozanov:2018kpk}.

In the spirit of \cite{ymmemo2,Jokela:2019apz} we shall assume the classical YM fields are those appearing in a nuclear wave function excited by a weak probe. Associating an axion with these phenomena is speculative. However, in the study of deep inelastic scattering on polarized hadrons a momentum structure $\epsilon^{\mn\alpha\beta}p_\mu q_\alpha \chi(p,q)$, analogous to that in \nr{sax}, naturally enters. This is due to the appearance of chiral triangle anomaly in polarized DIS, discussed, for example,  in \cite{Tarasov:2020cwl}. We suggest that it would be useful to study how the assumption of a particle-like axion state would fit in the framework of classical fields in large nuclei in the infinite momentum frame.

The action \nr{sax} leads to the equations of motion
\ba
 D_\mu F^\mn & = & J^\nu-\lambda \,\p_\mu\chi\,\tilde F^\mn ,\qquad D_\mu\,\tilde F^\mn=0  \la{eom1} \\
f^2\square \chi-V'(\chi) & = & \fra\lambda 4 F_\mn \,\tilde F^\mn \ . \la{eom2}
\ea
Defining the axion current
\be
 j_\rmi{ax}^\nu=-\p_\mu\chi\,\tilde F^\mn
\ee
we have automatically ($\chi$ is color singlet)
\be
 D_\nu j_\rmi{ax}^\nu =0 \ , \la{axcurrcons}
\ee
so that the current $J^\nu$ has to satisfy the condition
\be 
D_\nu J^\nu=0 \ .
\ee
We shall further split the current $J$ in components corresponding to a nucleus $A$ and probe $p$ moving along opposite light cones:
\be
J=J_A+j_p \ ,
\ee
where $J_A$ has only a $+$ and $j_p$ only a $-$ component.

To approximately solve the equations \nr{eom1} and \nr{eom2} we write
\be
 A_\mu = A_\mu^{(0)}+\lambda A_\mu^{(1)}+\ldots \equiv A_\mu +\lambda a_\mu+\ldots \qquad \chi = \chi_0+\lambda \chi_1+\ldots 
\ee
and iterate to first order in $\lambda$, treating $j_p^\nu$ as $\CO(\lambda^1)$. The intent is to include only the fluctuations caused by the axion, more generally there will be quantum fluctuations with different momentum spectra. Expanding in $\lambda$,
\ba
D_\mu(A +\lambda a)F^\mn( A +\lambda a)&=&
J_A^\nu+\lambda j_p^\nu + \lambda j_\rmi{ax}^\nu \la{full} \\
f^2\square (\chi_0+\lambda \chi_1)-V'(\chi_0+\lambda \chi_1) &=&\fra\lambda 4 F_\mn(A +\lambda a) \,\tilde F^\mn(A +\lambda a) \\
D_\nu(A +\lambda a)(J_A^\nu+\lambda j_p^\nu + \lambda j_\rmi{ax}^\nu) & = & 0 \ ,
\ea
leads to:

\vspace{4mm}
$\CO(\lambda^0)$ equations:
\ba
 D_\mu(A)F^\mn(A) & = & J_A^\nu \la{01}    \\
 f^2 \square \chi_0-V'(\chi_0) & = & 0  \la{02}\\
 D_\nu(A)J^\nu_A & = & 0 \ .\la{03}
\ea

$\CO(\lambda^1)$ equations (with one order $\lambda$ correction)
\ba
[D^2a^\nu - D^\nu D\cdot a+2ig F^\mn a_\mu]_c & = & -\p_\mu\chi_0\cdot \tilde F^\mn_c \la{11} \\
f^2\square\chi_1-\chi_1 V''(\chi_0) & = & F_\mn^a\tilde F^\mn_a+\lambda\tilde F^\mn  D_\mu a_\nu \la{12} \\
D_\nu(A)(j_p^\nu+j_\rmi{ax}^\nu)-ig a_\nu J_A^\nu & = & 0 \ . \la{13}
\ea
In \nr{11}-\nr{13} $D^\nu,F^\mn,\tilde F^\mn$ are adjoint representation matrices, $F=F^a T^a$, $T^a_{bc}=-if_{abc}$, evaluated at $A_\mu$, the solution of \nr{01}, $a_\mu^c$ is a color vector. The order $\lambda$ correction in \nr{12} is included since the leading term actually vanishes.

Note that $\lambda$ is here intended as a parameter counting axionic interactions. Perturbatively, there are also quantum fluctuations of order $g$, which are neglected here, the background field $A$ is taken to be
a purely classical YM field. Quantum effects will enter by integrating over an ensemble of color currents.

We shall now apply these equations for a process in which a weakly interacting probe $p$ moves along the $x^-$ axis and collides with a large nucleus $A$ moving along the $x^+$ axis, see Fig.~\ref{burst} (Left). On an event-by-event basis the collision excites the nucleus to an effective color field configuration together, as is assumed in this work, with a weak axionic configuration. We shall solve these configurations in order to check whether they can be represented in the framework of a memory effect.

We shall work here in the very high energy approximation of a Lorentz contracted infinitely thin nuclear sheet. The thickness parameter $\epsilon$ is taken to zero at the end of the computation. In usual discussions of the memory effect the coordinate
$x^\mu=(u,r,\theta,\phi)$, $u=t-r$, with the line element
\be
 ds^2=-du^2-2du\,dr +r^2(d\theta^2+\sin^2\theta d\phi^2)
\ee
is a natural one to use, for the angular part,  a more general metric $ds^2=h_{AB}\,d\theta^A d\theta^B$, $A,B=1,2$ on $S^2$. The relation to the light cone coordinates is simply $x_L=r\cos\theta$ with $\theta\to0$ 
so that the surface of $S^2$ is flattened. Then
\be
 \sqrt2 x^- = u+r-r\cos\theta\approx u
\ee
and the $t,x_L$ space-time diagram and the flat space Penrose diagram can be qualitatively related as in Fig.~\ref{burst}. In the Penrose diagram the null infinity is brought to a finite distance by a conformal transformation, in the $t,x_L$ space-time diagram the dominant field configuration is $x^+$ independent and the ``null infinity'' is at some large value of $x^+$.

\section{Leading order equations}\la{O0}

The {\cal O}$(\lambda^0)$ equation for the gauge field $A^\mu$ is, in color matrix$\times$color vector notation, 
\be
D_\mu F^\mn = J_A^\nu=\delta^{\nu +}\rho(x^-,\bfx) \ .
\la{YM}
\ee
Here $\rho$ is the color current of a nucleus $A$ moving in the $x^+$ direction in the infinite momentum frame. In the extreme thin sheet approximation one is tempted to write $\rho(x^-,\bfx)=\delta(x^-)\rho(\bfx)$, but when integrating over $x^-$ one should first take a finite range and then let the upper limit go to zero, see remarks around \nr{wrong}. Thus $\rho$ is concentrated in the range $0<x^-<\epsilon$, $\epsilon\to0$. It is also crucial for the following that there is no $x^+$ dependence, 
due to infinite time dilatation. Below we shall keep track of $x^+$ dependence, too. For a nucleus moving in the $x^+$ direction it is convenient to choose the light cone gauge\footnote{The logic of naming gauges is as follows. Since one $\pm$
component is fixed ($A^-=0$) the gauge is a light cone (LC) gauge. This comes in two variants, either as a longitudinal LC gauge, also called $A^+$ gauge ($A^+$ nonzero) or as a transverse LC gauge, also called $A^i$ gauge ($A^i$ nonzero).}
$A^-=-A_+=0$. Then a current with only $+$ component and no $x^+$ dependence automatically satisfies $D_\mu J^\mu=\p_+ J^+(x^-,\bfx)=0$, as required by \nr{YM}. 

Imposing first just the gauge condition $A^-=0$, $A^\mu=(A^+,0,A^2,A^3)$ the field tensor is, in the $(+,-,2,3)$ basis,
\ba
 F_\mn&=&\left( \begin{array}{cccc}0 & -\p_+ A^+ & \p_+A_2&\p_+A_3 \\
 \p_+ A^+ & 0 &F_{-2} & F_{-3}  \\ -\p_+A_2 & -F_{-2} &0 &F_{23}\\
-\p_+A_3 & -F_{-3} &-F_{23} &0
 \end{array} \right)
\nn &=&
\fr1{\sqrt2}\left( \begin{array}{cccc}0 & -\sqrt2E_L & E_2-B_3 &E_3+B_2 \\\sqrt2 E_L & 0 &E_2+B_3 & E_3-B_2 
\\- E_2+B_3 & -E_3-B_2 &0 & -\sqrt2B_L\\
-E_3-B_2 &-E_2+B_3 &\sqrt2 B_L &0
 \end{array} \right) \ .
\la{aminus}
\ea
Here the second form of $F_\mn$ records what the color electric and magnetic fields would be with the usual 3d associations $F_{0i}=E_i,\,\,F_{ij}=-\epsilon_{ijk}B_k$, where $i,j,k=1,2,3$, and remembering
that $x_1\equiv x_L$ is the longitudinal coordinate.

These explicit forms emphasize the strong effect of the approximation of $x^+$ independence, putting $\p_+=0$. Firstly, the longitudinal electric field vanishes, $E_L=0$. Secondly, the transverse electric and magnetic fields are related:
\be
 (E_2,E_3)=(B_3,-B_2) \ , \ E_i=\epsilon_{ij}B_j \ , \ B_i=-\epsilon_{ij}E_j \ , \   E_iB_i=0 \ .
\la{fields}
\ee
and orthogonal, $\tilde F^\mn F_\mn=0$ (see below). With no $x^+$ dependence, the only nonzero components of $F_\mn$ are $F_{ij}$ and $F_{-i}= F^{i+}=\p_-A_i+D_iA^+$ and the field tensor is
\be
 F_\mn=\left( \begin{array}{cccc}0 & 0 & 0 &0 \\ 0 & 0 &F_{-2} & F_{-3}  \\ 0 & -F_{-2} &0 &F_{23}\\
0 & -F_{-3} &-F_{23} &0
 \end{array} \right) \ .
\la{fmn}
\ee
For the dual tensor we have  ($\tilde F^{+i}=-\epsilon_{ij}F_{-j}$),
\be
\tilde F^\mn=
\fra12 \epsilon^{\mn\alpha\beta}F_{\alpha\beta}=
\left( \begin{array}{cccc}0 & F_{23} & -F_{-3} &F_{-2} \\ -F_{23} & 0 & \p_+A_3 & -\p_+A_2  \\ 
F_{-3} &- \p_+A_3  &0 & -\p_+A^+\\
-F_{-2} &\p_+A_2 & \p_+A^+ &0
 \end{array} \right)=
\sqrt2\left( \begin{array}{cccc}0 & 0 & B_2 &B_3 \\ 0 & 0 &0 &0  \\ -B_2 & 0 &0 &0\\
-B_3 &0&0 &0
 \end{array} \right)
\ee
with
\be
\tilde F^\mn F_\mn = \fra12\epsilon^{\mn\alpha\beta}F_\mn F_{\alpha\beta}=
4\,\Big[\p_+A_3\cdot F_{-2} -\p_+A_2\cdot F_{-3}- \p_+ A^+ \cdot F_{23}\Big].
\la{bdote}
\ee
The second form of the dual follows from $x^+$ independence and choosing $A^+$ gauge in which $F_{23}=0$ (see below). One sees concretely how the vanishing of $\tilde FF$ follows from $x^+$ independence.

We can now return to solving the YM equations \nr{YM} choosing $A^-=0$ and assuming $x^+$ independence. The $\nu=-$ component of \nr{YM}, $D_+F^{+-}+D_iF^{i-}=J_A^-=0$, is identically satisfied since the field tensor components vanish. The $+,i$ components of \nr{YM} are
\ba
 D_iF^{i+} & = & D_i\p_-A_i + D_iD_i A^+=\rho(x^-,\bfx) \la{am0} \\
 D_jF^{ji} & = & 0 \ .
\la{amj}
\ea
We remind that an equation $DF=\rho$ is short for a matrix equation $D_{ab}F_b=\rho_a$. Using the associations in \nr{fmn} we can equally write
\be
 D_iF^{i+}=\sqrt2 D_iE_i=\sqrt2 \epsilon_{ij}D_iB_j=\rho \ .
\ee

\begin{figure}[!t]
 \begin{center}
  \includegraphics[width=0.56\textwidth]{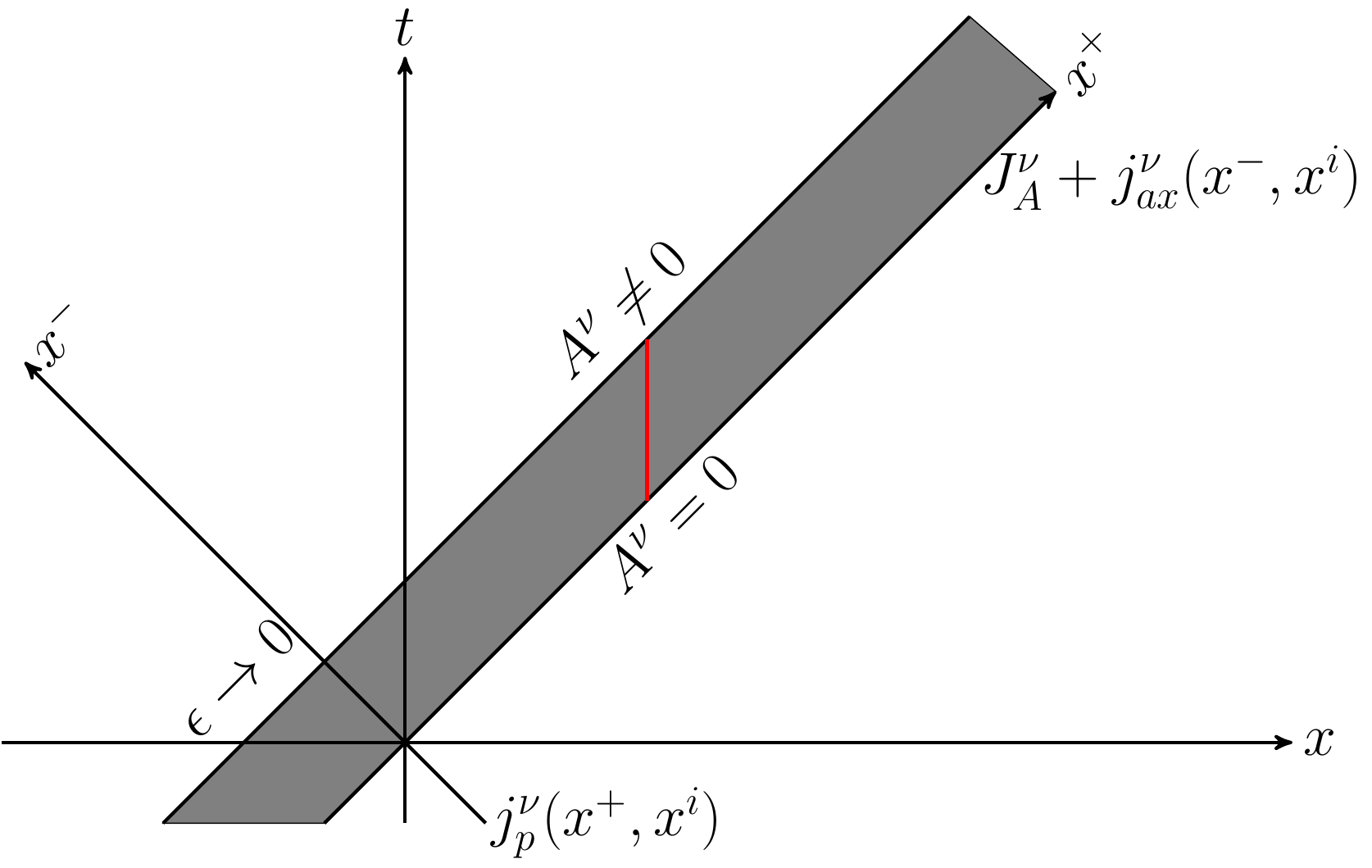}\hspace{1cm}
  \includegraphics[width=0.36\textwidth]{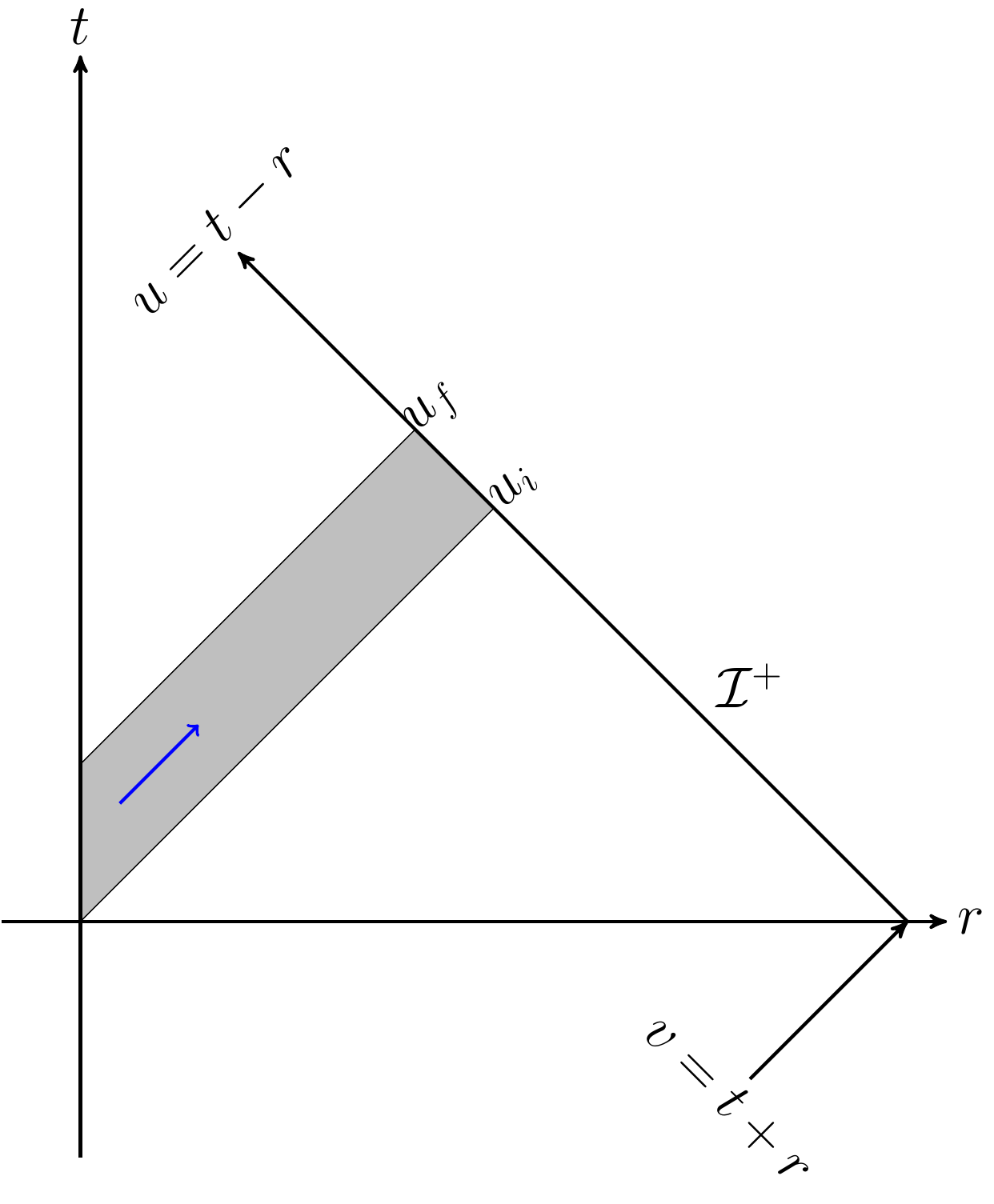}
 \end{center}
 \caption{\small  {\bf{Left:}} A weak probe $p$ moving along the $x^-$ axis (current $j_p^\nu$) collides with a large nucleus moving along the  $x^+$ axis (current $J_A^\nu$) spread over a distance $\epsilon$ in the $x^-$ direction. As discussed in Sections~\ref{O0}~and~\ref{O1}, a colored field configuration with an axionic component with the current $j_\rmi{ax}^\nu$ is excited in the strip $0<x^-<\epsilon$. 
Computations are carried out in the limit $\epsilon\to0$. Memory is the kick experienced by a test particle
(vertical red line, Section~\ref{memo}).  {\bf{Right:}} For comparison, a Penrose diagram presentation of the memory effect in electrodynamics. A radiator at $r=0$ sends a pulse of radiation to null infinity ${\cal I}^+$ during the time interval $u_i<u<u_f$. The time integrated pulse of transverse electric field gives a total momentum kick in \protect\nr{p} to a test charge at null infinity.
 }\la{burst}
\end{figure}

Further discussion of \nr{am0} and \nr{amj} splits naturally in two branches, one can find solutions with either the longitudinal light cone gauge (LLC) $A^i=0$, which we call $A^+$ gauge, this being the only non-zero component:
\ba
&&A^\mu =(\tilde A^+(x^-,x^i),0,0,0),\quad D_\mu =(\p_+,\p_-+ig\tilde A^+,\p_i) \nn
&&D^\mu =(-\p_--ig\tilde A^+,-\p_+,+\p_i),\quad D^2=\square-2ig\tilde A^+\p_+ \nn
&&F^{+-}=F^{-i}=0, \quad F_{-i}=\p_i \tilde A^+=\sqrt2 \tilde E_i=\epsilon_{ij}\sqrt2 \tilde B_j \ ,
\la{LLC}
\ea
or the transverse light cone gauge (TLC) $A^+=0$, which we call $A_i$ gauge, these being the only non-zero components:
\ba
&&A^\mu=(0,0,A^i(x^-,x^j)=\fra i g U\p_iU^\dagger),\quad \p_- U=igU\tilde A^+ \la{ai} \\
&&D_\mu =(\p_+,\p_-,\p_i-igA_i),\quad D^2=-2\p_+\p_-+D_iD_i \nn
&&F^{+-}=F^{-i}=0, \quad F_{-i}=\p_- A^i=\sqrt2 E_i=\epsilon_{ij}\sqrt2 B_j \ .
\la{TLC}
\ea
Whenever confusion might arise, quantities in $A^+$ gauge will be appended with a tilde. In these equations, $\tilde A^+$ is determined from the equation of motion \nr{am0} with $A^i=0$: 
\be
 D_\mu F^{\mu +}=D_iF^{i+}=\p_i\p_i \tilde A^+(x^-,x^i)=\tilde\rho(x^-,x^i)\ ,  \la{poisson}
\ee
{\emph{i.e.}}, by inverting a 2d transverse Poisson equation,
\be
 \tilde A^+(x^-,\bfx)=\int d^2y\,G(\bfx-\bfy)\tilde\rho(x^-,\bfy)=\fr1{2\pi}\int d^2y\log(|\bfx-\bfy|\Lambda) \tilde\rho(x^-,\bfy) \ ,   \la{A+green}
\ee
where $\Lambda$ is an IR cutoff parameter. A key role in the following is played by the adjoint matrix $U(x^-,\bfx)$ transforming from $A^+$ to $A_i$ gauge, {\emph{i.e.}}, transforming $\tilde A^+$ to zero. According to \nr{gt} this matrix $U$ has to satisfy
\be
 \p_-U^\dagger +ig\tilde A^+U^\dagger = D_-U^\dagger =0 \  ,  \la{D-U}
\ee
{\emph{i.e.}},
\be
 U(x^-,x^i)=P\exp\biggl[ig\int_0^{x^-} dy^- \tilde A^+(y^-,x^i)\biggr]U(0,x^i) \ ,
\la{U}
\ee
transforms $\tilde A^+$ to zero. Note that we define $U$ with $+ig$ in the exponent. Since
\be
\p_+U(x^-,\bfx)=0 \ ,
\ee
$A^-=0$ is intact and one transforms from the longitudinal to the transverse LC gauge. Since the first order matrix equation \nr{D-U} is homogeneous, its solution \nr{U} could be multiplied with an arbitrary matrix function $M(x^+,\bfx)$. The transverse field $A_i$ given in \nr{ai} is then generated and the field tensors transform, in matrix notation, as
\be
 F^{i+}=\p_-A^i=U\tilde F^{i+}U^\dagger=  U\p_i\tilde A^+U^\dagger \ .  \la{Ftrans}
\ee
In component form we can as well write, using \nr{uadj},
\be
\p_-A_{ia}=U_{ab}\, \p_i\tilde A^+_b \ , \ \p_i\tilde A^+_a=(U^{-1})_{ab}\,\p_-A_{ib}= U_{ba}\p_-A_{ib} \ . \la{etrafo}
\ee

As discussed above  $\rho$ and $A^+$ are confined in the range $0<x^-<\epsilon\to 0$, due to Lorentz contraction, and one might be tempted to insert $\delta(x^-)$ for the $x^-$ dependence. However, then $U$ in \nr{U} would be $\sim$
\be
 \exp[-igA^+(0,\bfx)]\theta(x^-)+\theta(-x^-)
\la{wrong}
\ee
and this form does not satisfy \nr{D-U}. One should keep the path ordered integral and only at the end take the range to zero.

The field configuration excited by a weak probe is thus very simple, just radiation-like mutually orthogonal color electric and magnetic fields. The situation is quite different in glasma, the state excited in a collision of two large systems \cite{Lappi:2006fp}. Then also longitudinal fields are excited.

Consider then the axion equation \nr{eom2} or its expanded versions \nr{02} and \nr{12}. Since the leading term for $F\tilde F$ vanishes, the equation to order $\lambda^0$ and to order $\lambda^1$ is simply the free scalar equation, 
\be
 \square \chi -m^2\chi =0 \ . 
\ee
The simplest approximation $V(\chi)=\fra12 f^2m^2\chi^2$, $m= $ axion mass, is used for the potential. Actually the axion will induce an order $\lambda^2$ inhomogeneous term $\fra{\lambda^2}{f^2}\p_+(a_i\sqrt2 B_i)$ to the RHS (see later Eq.\nr{newffd}).
Thus to $\CO(\lambda)$ the axion simply is a plane wave state $\chi_0(k)\, e^{ik\cdot x}$, $2k^+k^-=k_T^2+m^2$. Since the inhomogeneous term is of higher order the normalization is unknown; actually it is too much to expect that the normalization could be determined. If the axion is an exponential in time, $\sim e^{imt}$, then along $x^-=0$, just after the nuclear sheet, where we really need it,
\be
 \chi_0(mx^+,0,\bfx)=\chi_0(\bfx)e^{i\fra{m}{\sqrt2} x^+} \ .
\la{chimt}
\ee 
We will also need the combination
\be
 \fra1{\p_+}\p_+\chi_0=\chi_0( mx^+,0,\bfx)-\chi_0(0,0,\bfx)\equiv \bar\chi_0( mx^+,0,\bfx) \ , \la{invplus}
\ee
using the normalization \nr{invd+} of inverse $\p_+$.  This vanishes when $m\to0$ \cite{Hill:2015vma}. Of course, the $x^+$ integration constant in \nr{invd+} is basically unknown.

At this point one may also compare the situation in QCD and cosmological contexts, in view of Fig.~\ref{burst}. In the cosmological context \cite{Yoshida:2017fao} one has an emitter at $r=0$ which during a time interval $u_f-u_i$ sends a pulse to null infinity. Actually the emitted radiation is cosmological background radiation and null infinity is here, where the radiation is observed. In the course of its propagation the radiation passes through a (tentative) axionic dark matter background and this affects the polarization properties of the radiation so that not only $E$-mode radiation but also $B$-mode one is observed. This is a memory effect. There is only one universe, but the observed effect is still an average over many subsystems. In the QCD case, the field configuration is excited by a probe colliding with the nucleus. The transverse radiation potential $A_i$ grows within the shock wave $0<x^-<\epsilon$ essentially $\sim \theta_\epsilon(x^-,\bfx)$, so that the fields, derivatives of $A_i$ are 
$\sim \delta_\epsilon(x^-,\bfx)$ and can produce a finite result when integrated over $0<x^-<\epsilon\to0$. The required parity violation resides in the anomalous non-conservation of the axial current. It has been extensively discussed, in the form of the chiral magnetic effect, mainly in the central region of nucleus-nucleus collisions, less so in phenomena involving a single nuclear sheet (see, however, \cite{Tarasov:2020cwl}).

\section{Next-to-leading order equations}\la{O1}

Inserting the computed background field to \nr{11} we have the fluctuation equation
\be
D^2a^\nu - D^\nu D\cdot a+2ig F^\mn a_\mu =j^\nu_p+ j^\nu_\rmi{ax} \ ,
\ee
where
\be
j^\nu_p=\delta^{\nu -}\delta(x^+)\rho_p(\bfx),\quad j^\nu_\rmi{ax}= \epsilon_{ik}\p_k A^+(\delta^{\nu i}\p_+\chi_0-\delta^{\nu +}\p_i\chi_0) 
\la{currs}
\ee
and
\be
j^\nu_p+ j^\nu_\rmi{ax}=(\sqrt2 B_i\p_i\chi_0,\,\,\delta(x^+)\rho_p(\bfx), -\sqrt2 B_i\p_+\chi_0) \ , \la{curr}
\ee
where we introduced the magnetic field from \nr{fields}, $\sqrt2 B_i=-\epsilon_{ij}\p_j\tilde A^+$, and recall $\chi_0=\chi_0( mx^+,0,\bfx)$ in \nr{chimt}. We will discuss these equations in the $A^+$ gauge modified by the fluctuation field in the gauge $a^-=0$:
\be
 A^\mu+a^\mu=(A^+(x^-,\bfx)+a^+,0,a^i) \ ,
\ee
in which they have the explicit form ($D\cdot a=D_\mu a^\mu = \p_+a^+ +\p_i a^i$)
\ba
& & \nu = - \quad \p_+(\p_+a^+ +\p_i a^i) = j_p^-  \la{eqq1} \\
& & \nu = i \quad \square a_i-2igA^+\p_+a_i- \p_i(\p_+a^+ +\p_i a^i)  = j^i_\rmi{ax} \la{eqq2} \\
& & \nu = + \quad \square a^+-2igA^+\p_+a^++ ( \p_-+igA^+) (\p_+a^+ +\p_i a^i)+2ig\p_iA^+\cdot a_i  = j^+_\rmi{ax} \ .\la{eqq3}
\ea
We shall restore the tilde in notation for $A^+$ gauge when simultaneous quantities in the $A^i$ gauge start entering, after Eq.~\nr{tutudag}.

One is interested in solving these equations for the fluctuation field $a^\mu=(a^+,a^-=0, a^i)$ by integrating over the region depicted in Fig.~\ref{burst}, starting from vanishing values at $x^-=-\infty$ and then integrating in the direction of $x^-$. The main effect is what happens when crossing the nucleus, in the range $0<x^-<\epsilon\to0$. Note that this $x^-$ integration is in exact analogy when integrating over $u$ at large $r$ when computing the ED memory in $(u,r,\theta_A)$ coordinates.

Let us first check the current conservation condition \nr{13} explicitly. First, $a_\nu J_A^\nu=0$ since $J_A$ has only the $+$ component and $a_+=-a^-=0$. Contracting the current \nr{curr} with $D_\nu=(\p_+,D_-,\p_i)$ cancels the axionic terms (as should, according to \nr{axcurrcons}), but the condition
\be 
 D_-j_p^-=(\p_-+igA^+)j_p^-=(\p_-+igA^+)\delta(x^+)\rho(\bfx)=0   \la{jpcons}
\ee
remains. It is satisfied whenever $A^+=0$, but on the nuclear sheet at $0<x^-<\epsilon$ we are back to Eq.~\nr{D-U}, the collision with the nuclear sheet rotates the color of the probe by multiplying $j_p^-$ by the conjugate of the matrix $U^\dagger$ in \nr{U}. We thus have to write the probe current in the form
\be
 j_p^-=\delta(x^+)[\theta(x^-)U^\dagger(x^-,\bfx)\rho_p(\bfx)+\theta(-x^-)\rho_p(\bfx)] \ .
\ee

Returning to the fluctuation equations \nr{eqq1}-\nr{eqq3}, one first sees that the $\nu=-$ equation \nr{eqq1} can be integrated to
\be
 D\cdot a =\p_+a^+ +\p_i a^i={1\over\p_+}j_p^-= \theta(x^+)[\theta(x^-)U^\dagger(x^-,\bfx)\rho_p(\bfx)+\theta(-x^-)\rho_p(\bfx)] \ , \la{Daval}
\ee
using 
\be
 {1\over\p_+}f(x^+)=\int_0^{x^+} dy^+\,f(y^+) \ . \la{invd+}
\ee
The lower limit is at $x^+=0$ since that is when the collision takes place. In view of \nr{jpcons}, $D\cdot a$ also is covariantly conserved:
\be
 D_-(D\cdot a)=0\ , \la{Dacons}
\ee
as is expected of a ``time'' $x^-$ independent constraint.

Before the collision, at $x^-<0$, we have $A^+=0$ and all the fields are simple to solve. First,
\be
 (-2\p_+\p_-+\p_T^2)a_i-\theta(x^+)\theta(-x^-)\p_i\rho_p(\bfx)=0 \la{aieq1}
\ee
so that
\be
 a_i=\theta(x^+)\theta(-x^-){\p_i\over \p_T^2}\rho_p(\bfx)= \theta(x^+)\theta(-x^-)\int{d^2y\over2\pi}\,{x_i-y_i\over|\bfx-\bfy|^2}\rho_p(\bfy) \ , \ a^+=0 \ , \ D\cdot a = \p_i a^i \ . \la{initvals}
\ee
Actually for this solution $\p_+\p_- a_i\sim \delta(x^+)\delta(x^-)$ so that it only satisfies \nr{aieq1} away from the collision point $x^+=x^-=0$. The most general solution of \nr{Dacons} would contain $U^\dagger$ multiplied by a matrix function $M(x^+,\bfx)$, independent of $x^-$ \cite{Gelis:2008rw}.

Writing $\square$ explicitly and dividing by $-2\p_+$ the two last ones, Eqs. \nr{eqq2} and \nr{eqq3}, become 
\ba
&&\nu=i\quad \p_- a_i+igA^+a_i=-\fra1{2\p_+}\left(-\p_T^2 a_i+\p_i (D\cdot a)  +j^i_\rmi{ax}\right) \la{eqq2p} \\
&&\nu=+\quad  \p_- a^++igA^+a^+=-\fra1{2\p_+}\left(-\p_T^2 a^+ -2ig \p_iA^+\cdot a_i-D_-(D\cdot a) +j^+_\rmi{ax}\right) \ .\la{eqq3p}
\ea
We are particularly interested in integrating these across the nuclear sheet, $0<x^-<\epsilon$. Inserting what we learnt of $D\cdot a$ these in this range, and for $x^+>0$, are, in $A^+$ gauge,
\ba
 \nu=i\qquad D_-a_i=\p_- a_i+igA^+a_i &=&\fra1{2\p_+}\left(\p_T^2 a_i-\p_i (U^\dagger \rho_p) +\sqrt2 B_i\p_+\chi_0\right) \la{eqq2q}\\
\nu=+\quad D_- a^+= \p_- a^++igA^+a^+&=&\fra1{2\p_+}\left(\p_T^2 a^++2ig \p_iA^+ \cdot a_i-\sqrt2 B_i\p_i\chi_0\right) \ .\la{eqq3q}
\ea

As a check of the consistency of the equations \nr{eqq2q} and \nr{eqq3q} one may compute that their solutions indeed satisfy
\be
 D_- D\cdot a = D_-\p_+ a^+ + D_- \p_i a^i = 0 \ .
\ee

The equations \nr{eqq2q} and \nr{eqq3q} are 1st order inhomogeneous matrix equations which are solved by first solving the homogeneous equation and adding an inhomogeneous term. If $M,F$ are vectors and $A$ a matrix, the equation is of type
\be
 \p_xM(x)+A(x)M(x)=F(x) \ .
\ee
The homogeneous equation $\p_xM(x)+A(x)M(x)=0$ is solved by
\be 
 M_0(x)=P\exp\left[-\int_0^x dy\,A(y)\right]M_0(0)\equiv U^\dagger(x)M(0)
\ee
and the general solution is ($C$ is a constant)
\be
 M_a(x)=C\,U^\dagger_{ab}(x)M_b(0)+U^\dagger_{ab}(x)\int_0^x dy\,U_{bc}(y)F_c(y) \ .\la{gensol}
\ee

Consider now the equation \nr{eqq2q} for $a_i$. In it $A^+$ and 
$B_i$, as a spatial derivative of $A^+$, contain a $\delta(x^-)$ singularity, regulated by $\epsilon$. We expect that these singular terms dominate over the two transverse spatial derivative terms on the RHS. We shall therefore neglect these regular transverse terms (as was done in \cite{Gelis:2005pt} in an analogous computation). Note that their sum $\p_T^2 a_i-\p_i (U^\dagger \rho_p)$ vanishes if $a_i=\fra1{\p_T^2}\p_i(U^\dagger\rho_p)$. The equation then basically becomes an equation for $\p_+a_i$, but the RHS also depends on $\p_+\chi_0$. Dividing out $\p_+$ we have to use \nr{invplus} for the inverse. The equation for $a_i$ then becomes
\be
 \p_- a_i+ig\tilde A^+a_i=\fra1{\sqrt2}\tilde B_i \,\bar\chi_0(mx^+) \quad , \quad \sqrt2 \tilde B_i=  -\epsilon_{ij}\p_j \tilde A^+ \ .  \la{aieq}
\ee
To solve  this, we first need the homogeneous solution for $a_i$ with the initial condition \nr{initvals}:
\be 
 a_i^{(0)}(x^-,\bfx)= U^\dagger(x^-,\bfx) a_i(0,\bfx)\ , \  a_i(0,\bfx)=\fra{1}{\p_T^2}\p_i\rho_p(\bfx) \ . \la{aihom}
\ee
Then \nr{gensol} gives the transverse fluctuation field in $A^+$ gauge:
\ba
 a_i(x^+,x^-,\bfx) & = & U^\dagger(x^-,\bfx)\,a_i(0,\bfx) \nn 
 & & +\,\,U^\dagger(x^-,\bfx) \int_0^{x^-} dy^- U(y^-,\bfx)\fra1{\sqrt2}\tilde B_i(y^-,\bfx) \,\bar\chi_0(mx^+,y^-,\bfx) \ .\la{aisol1}
\ea
Here the upper limit $x^-$ is within the range $0<x^-<\epsilon$.

As the final step, we want this solution at the exit from the nuclear sheet, at $\epsilon\to0$. That the integral does not vanish in this limit follows from the fact that there effectively is a $\delta(x^-)$ singularity in $B_i$ on the nuclear sheet: a large background field $A_i\sim\theta(x^-)$ is created and $B_i$ is a derivative thereof. In the 2nd term, according to \nr{Ftrans}, the field derivatives are mathematically related by $\p_-A_i=U\p_i\tilde A^+ U^\dagger$ or, in terms of color vector components (see \nr{uadj}), by 
\be
 \p_- A^i_a=\fra1{N_c}\Tr(T_a\,U\, T_b\, U^\dagger)\p_i \tilde A^+_b =U_{ab}\,\p_i\tilde A^+_b\ . \la{tutudag}
\ee
The axion is effectively constant in the $y^-$ integration so that in the second term we can write (tildes are now restored)
\be
\int_0^{\epsilon} dy^- U(y^-,\bfx)\,\p_j \tilde A^+(y^-,\bfx)\bar\chi_0\approx\int_0^{\epsilon\to0}dy^-\p_- A_j(y^-,\bfx)\bar\chi_0 \approx A_j(\epsilon,\bfx)\bar\chi_0(x^+,0,\bfx) \ .
\la{eAchi}
\ee
Note that we are automatically lead to the adjoint color vector component
\be
 A_{jb}=\fra1{N_c}\Tr T_b A_j 
\ee
of the background field $A_j=\fra{i}{g}U\p_jU^\dagger$ in the $A_i$ gauge.

Thus, the transverse axion induced fluctuation field, in the $A^+$ gauge, at the exit from the nucleus is
\be
 \tilde a_{ia}(x^+,\epsilon,\bfx)=U^\dagger_{ab}(\epsilon,\bfx)  \Big[-\fra{1}{2} \epsilon_{ij}A_{jb}(\epsilon,\bfx)\bar\chi_0(mx^+,0,\bfx)+\fra1{\bf \p^2}\p_i\rho_{pb}(\bfx)\Big] \ .  \la{aisol}
\ee
We have written down the color components explicitly to emphasize the fact that one should take the color component $b$ of the vector $A_j$, not the matrix. 
In the second term the color index $b$ comes from the  color density $\rho_{pb}$ of the incident probe. This is rotated by the matrix $U^\dagger$ while crossing the nucleus. In the axionic first term the axion is color singlet and the color index $b$ is that of a gluonic transverse field $\sim A_{ib}$ excited from the background. Its color is further rotated by $U^\dagger$ while traversing the sheet.

When gauge transforming the background plus fluctuation system from the $A^+$ to the $A_i$ gauge some correction terms arise, relative to transforming only the background field \cite{Ayala:1995kg,Gelis:2008rw,Jeon:2013zga}. These terms are computed in Appendix~\ref{trafoLC}, but they can be neglected in the thin sheet limit. The result in the $A_i$ gauge is thus simple to obtain: just left multiply the $A^+$ gauge result by $U$. This cancels the matrix $U^\dagger$ in the right hand side and
\be
 a_{ia}(x^+,\epsilon,\bfx)= -\fra{1}{2} \epsilon_{ij}A_{ja}(\epsilon,\bfx)\bar\chi_0(mx^+,0,\bfx) +\fra1{\bf \p^2}\p_i\rho_{pa}(\bfx) \ . \la{aisolAi}
\ee
This result implies that the axion has induced an $x^+$ dependence 
\be
 \p_+ a_{ia}=-\fra12\epsilon_{ij}\p_+(A_{ja}\bar\chi_0)=
i\fra{m}{\sqrt2}\chi_0(mx^+)(-\fra12\epsilon_{ij}A_{ja}) \la{xpdep}
\ee
to the transverse field. There is also an interesting property of probe-nuclear sheet interactions contained in the second term on the RHS of \nr{aisolAi}: the non-axionic transverse fluctuation is not at all affected by the sheet \cite{Kajantie:2019nse}. The last term is simply the vacuum solution \nr{initvals} before the collision.

The $\tilde a^+$ fluctuation (actually one only needs its derivative $\p_+\tilde a^+$) should now be solved from the equation \nr{eqq3q} 
\be
 \p_- \tilde a^++ig\tilde A^+\tilde a^+=\fra1{\p_+}\left[+ig \p_i\tilde A^+ \cdot \tilde a_i-\fra{1}{\sqrt2} \tilde B_i\p_i\chi_0\right] \ , \la{ap1}
\ee
where the non-singular term $\p_T^2\tilde a^+$ has been neglected. The solution can be directly written down from the general formula \nr{gensol} noting that since the initial condition is $a^+(0)=0$ (Eq.~\nr{initvals}), there is no homogeneous solution. Inserting $\tilde a_i$ from \nr{aisol} one to begin with has a non-axionic contribution from the homogeneous term of $\tilde a_i$. This, written for $\p_+\tilde a^+$ is, 
\be
 \p_+\tilde a^+=U^\dagger(x^-)\int_0^{x^-} dy^- \left(U(y^-)ig\p_i \tilde A^+\cdot U^\dagger(y^-)  a_i(0) \right)=U^\dagger\,igA_ia_i(0)=-\p_iU^\dagger a_i(0) \ ,
\ee
in agreement with \cite{Gelis:2005pt} (there $U^\dagger$ is defined as $U$). The relevant new axionic terms come from the inhomogeneous axionic term in \nr{aisol} and the last term in \nr{ap1}. The full result requires one more $y^-$ integral and is, with color indices,
\ba 
 \p_+ \tilde a^+_a & = & U^\dagger_{ab}\left[igA_{ibe} a_{ie}(0)- \fra12 \epsilon_{ij}ig\int_0^\epsilon dy^-\p_-A_{ibe} A_{je}\bar\chi_0+\fra12\epsilon_{ij}A_{jb}\p_i\chi_0 \right]\nn
 & = & U^\dagger_{ab}\left[igA_{ibe} a_{ie}(0)-\fra12\epsilon_{ij}(\p_iA_{jb})\bar\chi_0 +\fra12\epsilon_{ij}A_{jb}\p_i\chi_0 \right] \ , \la{apsol}
\ea
where the arguments $x^-=\epsilon,\,\,y^-,\bfy$ are omitted, $\chi_0=\chi_0(mx^+,0,\bfx)$ is as given in \nr{chimt}, $\bar\chi_0$ has the value at $x^+=0$ subtracted. 
By using symmetries the $y^-$ integral simply is $\fra12 A_{ibe}A_{je}$ so that the whole
middle term is $-\fra14 \epsilon_{ij}ig A_{ibe} A_{je}\bar\chi_0$ ($A_i$ appears both as a matrix and a vector here). However, one further has $igA_{ibe}A_{je}=ig[A_i,A_j]_b=\p_iA_{jb}-\p_jA_{ib}$ 
since the background solution is $F_{ij}=0$. Thus the middle term reduces to 
$-\fra14 \epsilon_{ij}ig A_{ibe} A_{je}\bar\chi_0=-\fra12\epsilon_{ij}\p_iA_{jb}\bar\chi_0$. 
With a different sign the axionic terms would combine to $\pm \fra12\epsilon_{ij}\p_i(A_{jb}\bar\chi_0)$.
Compare also with \nr{xpdep} for $\p_+a_i$. Note also how the sources of the two terms
are different, the middle term comes from the interaction with the transverse fluctuation, the
last term from the interaction of the $a^+$ fluctuation with the background field, see \nr{ap1}.

In summary, at the exit from the crossing of the nuclear sheet, in the $A_i$ gauge, the total transverse field
and the $x^+$ derivative of the longitudinal gluon field are given by
\be
 A_{ia}(\epsilon,\bfx)+a_{ia}(x^+,\epsilon,\bfx)=\Big(\delta_{ij}-\fra1{2}\epsilon_{ij}\bar\chi_0(mx^+,0,\bfx)\Big) A_{ja}(\epsilon,\bfx)+a_{ia}(0,\bfx) \ .\la{aisolsumm}
\ee
\be
 \p_+a^+_b(x^+,\epsilon,\bfx)=\fra12\epsilon_{ij}A_{jb}\,\p_i\chi_0(mx^+,0,\bfx)-\fra12\epsilon_{ij}(\p_iA_{jb})\bar\chi_0+igA_{ibe} a_{ie}(0)\ . \la{Elong}
\ee
In the $A^+$ gauge the fluctuation fields are in \nr{aisol} and \nr{apsol}. The corresponding field
tensors, in the $+,-,2,3$ basis, are, in the $A_i$ gauge
\be
 F_\mn(A+a)=\left( \begin{array}{ccc}0 & -\p_+ a^+ & \p_+a_i \\
 \p_+ a^+  &0 & \p_-(A_i+a_i)+ D_ia^+ \\ 
-\p_+a_i & {\rm antis} &  D_i a_j-D_j a_i
 \end{array} \right) \ . \la{newffd}
\ee
or in the $A^+$ gauge
\be
 F_\mn(\tilde A+\tilde a)=\left( \begin{array}{ccc}0 & -\p_+ \tilde a^+ & \p_+\tilde a_i \\
 \p_+ \tilde a^+  &0 & \p_i (\tilde A^++\tilde a^+)+ D_-\tilde a^i \\ 
-\p_+\tilde a_i & {\rm antis} &  \p_i \tilde a_j-\p_j \tilde a_i
 \end{array} \right) \ , \la{newffdAp}
\ee
All the fields are evaluated at $x^-=\epsilon\to0$, just after crossing the thin nuclear sheet, $a_{ia}(0,\bfx)=  \p_T^{-2}\p_i\rho_{pa}(\bfx)$ (Eq. \nr{aihom}), $\bar\chi_0$ is in \nr{invplus}.

Axion induced effects as as follows.  The large transverse gauge field $A_i$ is 
corrected by a perpendicular vector, the $-\fra12\epsilon_{ij}A_j\bar\chi_0$ term in \nr{aisolsumm}. 
This implies that the length of the color vector $A_{ia}$ is only changed
by a very small color independent amount
\be
 A_{ia}A_{ia}\to (1+\fra14\bar\chi_0^2(mx^+,\epsilon,\bfx)A_{ia}A_{ia}\approx A_{ia}A_{ia}\ . \la{length}
\ee 
This correction may decouple in the limit $m\to0$. The $a^+$ fluctuation is corrected by a term with a
very similar structure in \nr{Elong}.

The corrections to the color electric and magnetic fields induced by the axion can be read from 
\nr{newffd},\nr{newffdAp} together with \nr{aminus}:
\be
\sqrt2 E_i=\p_-A_i+\p_+a_i+\p_-a_i+D_ia^+,\quad B_i=-\epsilon_{ij}(E_j-\sqrt2\p_+a_j),
\ee
\be
\sqrt2 E_L=\p_+a^+,\quad B_L=-D_i a_j+D_j a_i.
\ee
In particular, a nonzero $F\tilde F$ is induced:
\be
 \fra14F_\mn \tilde F^\mn = -\epsilon_{ij}\p_+ \tilde a_i \,\p_j\tilde A^+ =\p_+\tilde a_i\,\sqrt2\tilde B_i = -\epsilon_{ij}\p_+  a_i \,\p_- A_j =\p_+a_i\,\sqrt2 B_i
\ee
in the two gauges. These are analogous to writing ${\bf E}\cdot{\bf B}=\p_t{\bf A}\cdot {\bf B}$ (3d vectors) in electrodynamics. One sees how $x^+$ dependence of the transverse fluctuation leads to a nonzero $F\tilde F$. The term $\p_+a^+$ in \nr{bdote} does not contribute since it is multiplied by $F_{23}$ which also is of first order. Using \nr{xpdep} we can further write
\be
\fra14 F_\mn^a \tilde F^\mn_a = \p_+ a_{ia}\cdot\sqrt2 B_{ia}  =-\fra12i\,m\chi_0\,\epsilon_{ij}A_{ja}B_{ia} =\fra12 i\,m\chi_0(mx^+)A_{ia}E_{ia} \ .\la{eqq5}
\ee
Remember that here $A_i,E_i,B_i$ are independent of $x^+$. This is valid as it stands in $A_i$ gauge but going over to $A^+$ gauge, where there is no background $A_i$ field, one must transform $A_{ja}$ in \nr{eqq5} to $U^\dagger_{ab}A_{jb}=(U^\dagger A_jU)_a =-\fra{i}{g}(U^\dagger\p_jU)_a$. This is in agreement with $\tilde A_{ia}=(U^\dagger A_iU)_a +\fra{i}{g}(U^\dagger\p_iU)_a=0$.

Many of the qualitatively important effects induced by the axion seem to come from the $x^+$ dependence of the fluctuations. The large background fields were independent of $x^+$.

\section{Memory}\la{memo}

In the set-up of Fig.~\ref{burst} the memory of YM radiation is the permanent effect this radiation burst has on some property of a test particle crossed by the burst. Without the axion the simplest type of memory \cite{Pate:2017vwa,ymmemo2,Jokela:2019apz,Campoleoni:2019ptc} is the transverse momentum change of the test particle, caused by the transverse electric field of the burst. We set out to study how the introduction of an axion-like particle would modify this pattern. We have now computed the color fields in the infinitesimally thin nuclear sheet approximation and the response of a test particle to these fields can, in principle, be computed from Wong's equations \cite{wong}.

Wong's equations give the motion $x^\mu=x^\mu(\tau)$ of a particle of mass $M$ with an adjoint color vector $Q_a$ in a given background field. Defining first 
\be
 p^\mu=Mu^\mu=M{dx^\mu\over d\tau}
\ee
they are ($Q\cdot F\equiv Q_aF_a$)
\be
 {dp^\mu\over d\tau}=gQ\cdot F^{\mu\nu}{dx_\nu\over d\tau}\ ,  \ {dQ^a\over d\tau}=-gf_{abc}u^\mu A^b_\mu Q^c \ .\la{eom}
\ee
Note that the equation for $p_\mu$ explicitly conserves the mass shell condition $p_\mu p^\mu=-2p^+p^-+p_i^2=-M^2$. The color equation expresses its covariant conservation: In matrix form ($u^\mu\p_\mu=\p_\tau$)
\be
 \dot Q-igu^\mu A_\mu Q=u^\mu(\p_\mu-igA_\mu)Q =u^\mu D_\mu Q=0 \ .
\ee
The $-,\,i,\,+$ components of the equations of motion are
\ba
 M{dp^-\over d\tau} & = &-gQ\cdot (F_{+-}p^-+F_{+i}\,p^i)=gQ\cdot (p^-\p_+ a^+-p^i \p_+ a^i) \ . \la{mshell1}  \\
 M{dp^i\over d\tau} & = & gQ\cdot \left(p^-F_{i-}+p^+F_{i+}+p^jF_{ij}\right) \nn
                    & = & gQ\cdot\left[-p^-\left(\p_-(A_i+a_i)+D_ia^+\right)-p^+\p_+ a_i+p^j(D_ia_j-D_ja_i)\right] \la{mshell2} \\
                    & = & g\tilde Q\cdot\left[-p^-\left(\p_i(\tilde A^++\tilde a^+)+D_-\tilde a^i\right) -p^+\p_+ \tilde a_i+p^j(\p_i\tilde a_j-\p_j\tilde a_i)\right] \la{mshell3} \\
p^+  & = & {p_ip_i+M^2\over 2p^-} \ .
\la{mshell}
\ea
We know the fields from the front of the nucleus at $x^-=0$ to its tail end at $x^-=\epsilon$ and we should compute the cumulative effect integrated over the nuclear sheet on a test particle starting at $x^-=0$ with some initial velocity $u^-(0)$, the fate of the red line in Fig.\ref{burst}.
All the fields are constructed on the basis of the large background transverse field $A_i(x^-,\bfx)$ together with the axion $\chi_0(x^+,x^-,\bfx)$. There is no reason to expect any strong variation as a function of $x^-$ in the axion wave function so that one can as well set $x^-=0$ there. The transverse field $A_i(x^-,\bfx)$ grows rapidly across the nuclear sheet, behaves $\sim \theta_\epsilon(x^-)$. There also the integral over the burst, over the range $0<x^-<\epsilon$ will produce something of the order of $\epsilon$. However, there is one term containing a singularity in the range of integration, the $x^-$ derivative $\p_-(A_i+a_i)$ in $dp^i/d\tau$, goes $\sim\delta(x^-)$ and produces a finite result in the limit $\epsilon\to0$. This feeds itself further into the behavior of $p^+$. Similarly, the $A^+$ gauge equation \nr{mshell3} has a $\delta(x^-)$ singularity in $\tilde A^+$ ; this will be discussed in Appendix~\ref{aplusgauge}.

We thus conclude that in the thin sheet limit we can concentrate on the $\p_-(A_i+a_i)$ term in $dp^i/d\tau$, the rest will produce $\CO(\epsilon)$ effects. However, this is a limit and in serious modeling the $\CO(\epsilon)$ effects should be quantitatively studied. There are also 
further $\CO(\epsilon)$ effects, like the one coming from careful gauge transformation between $A^+$ and $A_i$ gauges when also fluctuations are included, studied in Appendix~\ref{trafoLC}. Quantitative conclusions
are only possible by numerical means. Solving Wong's equations numerically has been extensively studied \cite{Moore:1997sn,Dumitru:2006pz,Li:2020uhl}.

Consider then Eq.~\nr{mshell1} for $M\,dp^-/d\tau$. The RHS is, from the point of view of the axion, particularly interesting since it is entirely induced by the $x^+$ dependence of the axionic fluctuation. 
Its coefficients, given in \nr{xpdep} and \nr{Elong}, have a reasonably simple structure, but the equation is
not obviously integrable. 
It is nevertheless non-singular and produces negligible $\CO(\epsilon)$ effects. We thus have
\be
 {dp^-\over d\tau}=0\quad \Rightarrow\quad p^-=Mu^-=M{dx^-\over d\tau}={\rm constant}  \quad\Rightarrow\quad x^-(\tau)={p^-\over M}\tau \ .
\ee
Of course, it will be a very interesting problem to ultimately sort out how the now neglected
coefficients in \nr{xpdep} and \nr{Elong} affect the constancy of $p^-$, but this requires a good
numerical control of the fields as well as a better knowledge of the axion wave function.

Assume then that the test particle is initially is at rest, $p^-=M/\sqrt2$.
The equations of motion conserve the mass shell condition so that
all the time during motion across the sheet
\be
 \sqrt2 p^-=E-p_L =\sqrt{p_L^2+p_T^2+M^2}-p_L=M_Te^{-y} = M   \la{plpt}
\ee 
From this one can solve
\be
p_L={p_T^2\over 2M}, \quad y=\log{\sqrt{p_T^2+M^2}\over M} \ .
\ee
so that the momentum of the test particle is
(in $(E,p_L,p_2,p_3)$ coordinates)
\be
p^\mu=\left({p_T^2\over 2M}+M,{p_T^2\over 2M},p_i\right) \ ,
\la{pmuT}
\ee
$p_ip_i=p_T^2$. Computing $p_i$ as a function of time, this 
gives the fate of the red line in Fig.\ref{burst}.
Passage through the sheet develops some $p_T$ and, associated with thus some
$p_L$. This is negligible in the non-relativistic limit, $p_T\ll M$. 
How this affects the U(1) memory analogy is discussed later after Eq.~\nr{longmem}.

Thus the primary quantity is transverse motion, the rest follows from it.
The equation for $x_i(\tau)$ is
\be
 \dot p_i=M\ddot x_i(\tau)= -g Q\cdot E_i(\tau,\bfx), \la{transeq}
\ee
where the color electric field is $E_i=F_{-i}/\sqrt2$ as given by \nr{newffdAp} or \nr{newffd}. Solving from here $x_i(\tau)$ one gets $x_L(\tau)$ by integrating $\dot x_L=\fra12 \dot x_i\dot x_i$ and finally (from $E=p_L+M$) $x^0(\tau)=\tau + x_L(\tau)$.

To do the first integral over $\tau$ or $x^-$ it is simplest to use the $A_i$ gauge since then $\sqrt2 E_i=\p_- A_i$ and one integral can be immediately carried out. Including just the $x^-$ derivative term in \nr{mshell2} the $p^i$ equation 
is simply
\be
 \p_- p_i(x^-)=-gQ_aF^{i+}_a(A+a)=-gQ_a\p_-(A_a^i+a_a^i)\ . \la{derpi}
\ee
In the $A_i$ gauge $D_-Q_a=\p_-Q_a=0$ and the color does not rotate. This is a key property of
the gauge choice since then 
we can immediately integrate over $x^-$. Choosing the initial value $p_i(0)=0$ and 
taking the transverse fluctuation from \nr{aisolsumm} (the weak probe initial field $a_{ia}(0,\bfx)$ is inessential and can be neglected), the final result for the transverse kick is\footnote{Of course, 
one can as well use the $A^+$ gauge, {\emph{i.e.}}, start from \nr{mshell3}. This is done in Appendix~\ref{aplusgauge}. For the consistency of the approach it is important that a gauge invariant answer is obtained.}
\be
 p_i(x^+,\bfx)=-gQ_a\left(A_a^i+a_a^i\right)=-g\left[\delta_{ij}-\fra12 \epsilon_{ij}\bar\chi_0(mx^+,0,\bfx)\right]Q_aA_{ja}(\bfx)  \ . \la{p}
\ee

Thus, in analogy with \cite{Yoshida:2017fao}, there is a new parity breaking mode. Both $A_{ia}$ and $p_i$ have the same parity ($P-$) and since $\chi_0$ is pseudoscalar, the $\epsilon_{ij}$-term has opposite parity. For given colors \nr{p} is a definite prediction for the transverse kick on an event-by-event basis, for one element of the nuclear color densities, given concretely in \nr{kickmag}. Note that $A_{ia}$ as an adjoint color vector is real. Colors and momentum dependencies remain unspecified, though, and in this sense this result may be mathematically correct but is unphysical.

Geometrically, the axion dependent correction in \nr{p} is a small perpendicular addition to the 2d vector $A_i$. Thus to first order, as already discussed in Eq.\nr{length}, the length of the vector $A_i$ is unchanged.

The result \nr{p} should now be (complex) squared and averaged over an ensemble of color densities. For a Gaussian ensemble, see Eq.~\nr{ensemble}. In this heavy ion collision analogue model, this averaging is concretely carried out in Appendix~\ref{aplusgauge}. One has to evaluate expectation values of the type (one takes different points $x=(x^+,\bfx),\,y=(y^+,\bfy)$ since there will be a logarithmic divergence when $\bfy\to\bfx$)
\ba
 &&\langle p_i(x^+,\bfx)p_i(y^+,\bfy)\rangle \la{pixpiy} \\
 &&\hspace{-8mm}=g^2\left\langle\Bigl[\delta_{jk}-\fra12\epsilon_{jk}\left(\bar\chi_0(mx^+,0,\bfx)-\bar\chi_0(my^+,0,\bfy)\right)+\fra14\delta_{jk}\bar\chi_0(x)\bar\chi_0(y)\Bigr] Q_aQ_b A_{ja}(\bfx)A_{kb}(\bfy)\right\rangle_\rho\la{pp}\nonumber \\
&&\underset{y\to x}{\longrightarrow} g^2 \Bigl[1+\fra14\bar\chi_0^2(mx^+,0,\bfx)\Bigr]\langle\Tr\left[A_i(\bfx)A_i(\bfy)\right]\rangle_\rho \ ,
\ea
where we on the 2nd line have taken $y\to x$ in the axionic factor and used the fact that the field expectation values are diagonal in color so that one can replace $Q_aQ_a\to C_A=N_c$, the adjoint Casimir, and write the color sum as a trace. Eq.~\nr{pp}
shows that unless there are some special effects in the $x^+$ direction, the first order axionic correction to the displacement memory vanishes. Note that this vanishing happens on the tree level, even before computing the expectation value. There will be corrections to next order. Physically, when traversing the thin nuclear sheet the color neutral axion had to pick up an adjoint color vector and there was only one available, $A_i$.

To relate the result to phenomenology, the leading memory term was evaluated in \cite{Jokela:2019apz}:
\be
g^2\langle\Tr\left[A_i(\bfx)A_i(\bfy)\right]\rangle_\rho =
 \lim_{\bfy\to\bfx}\langle p_i(\bfx)p_i(\bfy) \rangle={1\over\pi} Q_s^2 \log{Q_s\over\Lambda} \la{memfin}
\ee
where $Q_s$ is a saturation scale (of the order of 2 GeV) and $\Lambda$ (of the order of $\Lambda_\rmi{QCD}\approx m_\pi$) regulates the divergence at $\bfy\to\bfx$. The derivation was carried out in the $A_i$ gauge and required a computation of the expectation value of a string of $U$ matrices. This has more accurately been carried out in \cite{Lappi:2017skr}. As shown by \cite{Kolbe:2020hem}, it is simplest to use the $A^+$ gauge, then one gets not only the expectation value but the entire distribution, see Appendix~\ref{aplusgauge}. The results coincide which shows the consistency of the scheme.

Further, this analogue model predicts that in addition to the transverse displacement memory there is a longitudinal memory due to \nr{plpt}:
\be
 \langle p_L\rangle_\rho = \left\langle {p_T^2\over 2M} \right\rangle_\rho \ , \la{longmem}
\ee
where $M$ is the mass of the test particle. This effect is there already on event-by-event basis,
see \nr{pmuT}, and survives averaging over an ensemble of collisions.
The appearance of the infrared sensitive quantity $M$ indicates
that the longitudinal component of the memory is not as controllable as the transverse one.

That the longitudinal memory does not appear in usual discussions 
\cite{bieriED},\cite{Mao:2017wvx,Hamada:2018cjj,Hamada:2017atr},\cite{Campoleoni:2019ptc}
is due to the fact that this analogue model is inherently relativistic with equal
$E_i,\, B_i$ while in the usual
Lorentz factor ${\bf E}+\fra{1}{c}\bfv\times c{\bf B}$ with ${\bf E}\sim c{\bf B}$ 
(3-vectors) the magnetic field term is negligible at non-relativistic velocities. It is this term which
produces longitudinal motion.

\begin{figure}[!ht]
 \begin{center}
  \includegraphics[width=0.3\textwidth]{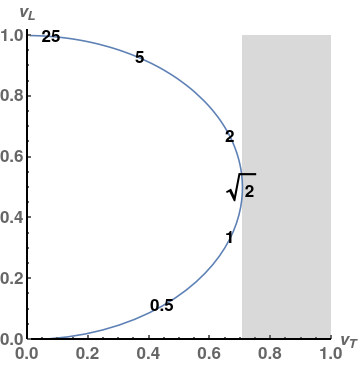}
 \end{center}
 \caption{The velocities in Eq. \protect\nr{vels} plotted on the $(v_T,v_L)$-plane as functions of $p_T/M$.
 }\la{ellips}
\end{figure}

To analyse this from another angle, note that from \nr{pmuT} the velocities are
\be
v_T={2Mp_T\over p_T^2+2M^2},\quad v_L={p_T^2\over p_T^2+2M^2},\quad 
v_L(1-v_L) =\fra12 v_T^2 \ .
\la{vels}
\ee
These velocities form an ellipse plotted on the $(v_T,\,v_L)$-plane in Fig.~\ref{ellips}.
In the nonrelativistic limit $p_T\ll M$, we have $v_T=p_T/M,\,\,v_L=p_T^2/(2M^2)=\fra12 v_T^2$
so that the longitudinal effect is negligible. Increasing $p_T$ $v_T$ grows and reaches its maximum
value $v_T=1/\sqrt2$ at $p_T=M\sqrt2,\,\,v_L=\fra12$. Increasing $p_T$ further, $v_T$ decreases 
and finally vanishes as $2M/p_T$ when $v_L\to1$. 
It may seem paradoxical that the transverse velocity decreases
in the large momentum limit; this of course is due to energy increasing.

In discussions of U(1) memory \cite{bieriED} there are two types of memory, an ordinary memory caused by a radial electric field (sourced by a collection of
charged particles which end up in the future timelike infinity, not in null infinity) 
and a null memory caused by transverse electric and magnetic fields (sourced by massless charged particles going to null infinity). In this analogue model there are
longitudinal fields sourced by the axion, but all sources go to null infinity. In this sense the model
is the analogue of the null memory only, caused by transverse fields.
Longitudinal memory in \nr{longmem} is a relativistic effect in the observation of null memory. 
Also longitudinal fields are induced by the axion but
we are so far unable to compute their effect. Longitudinal fields enter in the central region of nucleus-nucleus collisions \cite{Lappi:2006fp,Kharzeev:2015kna,Lappi:2017skr} or in phenomenological analyses of $\eta^\prime$ production \cite{jeonetapr}.

\section{Conclusions}\label{sec:conclusions}

This article was motivated by a cosmological study \cite{Yoshida:2017fao} in which the effect of a cosmological axion background on electromagnetic memory was computed. This gave a motivation to ask how an axion-like particle, called axion,  coupling to QCD matter would affect the Yang-Mills radiation memory in \cite{Pate:2017vwa,ymmemo2,Jokela:2019apz,Campoleoni:2019ptc}. Operationally, answering this required sorting out how an axion could coexist with a CGC, Color Glass Condensate.

Another physical way of expressing the problem is as follows. Assume there is a speculative axion-like
degree of freedom in QCD matter, interacting with QCD as axions are expected to do. How does
it affect the motion of a test quark passing through a large nucleus, in the infinite momentum frame?

This problem has been studied in the setting of a weak probe, a proton, an offshell photon, 
exciting the wave function of a single nucleus. Of course, there is no observational evidence of this
type of dynamics, Subtle analyses of spin effects in deep inelastic scattering may, nevertheless,
lead to some related effects \cite{Tarasov:2020cwl}.

The effect of the axion on the CGC in the form of axion-induced fluctuations of the 
gauge potentials in the CGC has been computed in the thin sheet limit. These effects can be
measured by the effect of the nuclear sheet on the motion of a test quark.
There is a clear parity violating memory effect on the classical event-by-event level, but summing over an ensemble of events, the axionic effect averages out. The treatment is inherently relativistic and the memory,
in addition to the usual transverse one \cite{bieriED}, has a longitudinal component also.
In any case, the memory here is an analogue of the null memory only, all the charges reach
null infinity.

There are lots of finite width effects to modify the averaging out of the axionic signal,  
with bigger width there is more space and time for interesting phenomena to take place. However,
these are accessible only by numerical computations, equations for which 
have been written down. Proceeding further would be an entirely new project.

Technically, the work presented here is largely based on the methods and approximations
developed in \cite{Gelis:2005pt}. There also gluon production in the process was computed. The
same can be done here, too, on the amplitude level, but averaging over an ensemble would need
computing very complicated correlators. The level of complication is set by a related computation of
$\eta'$ production \cite{jeonetapr}.

One knows of the axion little beyond the assumed free pseudoscalar massive field equation. It has one interesting effect, through its mass dependence it induces $x^+$, ``time'', dependence to the CGC. Inherently, due to time dilatation, the CGC is $x^+$ independent. Altogether, of course the whole appearance of the axion is speculative, but maybe this is a useful theoretical exercise anyway.

\vspace{0.8cm}
 {\bf Acknowledgements} We thank F.~Gelis, I.~Kolbe, T.~Lappi, and R.~Paatelainen for discussions on Color Glass Condensate and L.~Bieri, D.~Garfinkle, C.~Heissenberg, D.~Nichols, and B.~Oblak for discussions on gravitational radiation memory.
N.~J. and M.~S. have been supported in part by the Academy of Finland grant no. 1322307. M.~S. is also supported by the Finnish Cultural Foundation.

\appendix
\vskip 1cm

\renewcommand{\theequation}{\rm{A}.\arabic{equation}}
\setcounter{equation}{0}
\medskip
\section{Transformation between gauges}\la{trafoLC}

Let us work out in some detail the gauge transformation from the gauge $A^\mu=(\tilde A^++\tilde a^+,\,0,\,\tilde a_i)$ to the gauge $(0,0,A_i+a_i)$ \cite{Ayala:1995kg,Gelis:2008rw,Jeon:2013zga}. Transforming the $+$ component to zero requires
\be
 A^++a^+=0=\bar U(\tilde A^++\tilde a^+)\bar U^\dagger-\fra{i}{g} \bar U\p_-\bar U^\dagger
\ee
or 
\be 
 \p_-\bar U^\dagger = -ig(\tilde A^++\tilde a^+)\bar U^\dagger \ .
\ee
Here the transformation matrix $\bar U$ is expected to be close to  $U=e^{i\theta}$:
\be
 \bar U= e^{i(\theta + g\delta\theta)}=U(1+ig\delta\theta)\ .
\ee
Inserting this, expanding in $g$, using the leading relation $\p_- U^\dagger = -ig\tilde A^+ U^\dagger$, one finds that $\delta\theta$ is determined from the equation
\be
 D_-(\tilde A^+)\delta\theta =(\p_-+ig\tilde A^+)\delta\theta = \tilde a^+ \ ,
\ee
from which one solves
\be
 \delta\theta=U^\dagger \delta\theta(0)+U^\dagger(x^-,\bfx)\int_0^{x^-}dy^-U(y^-,\bfx)\tilde  a^+(y^-,\bfx) \ . \la{dth}
\ee 
Since $\tilde A^+$ vanishes for $x^-\le0$ one expects $\delta\theta(0)=0$ here. The other components are transformed to
\ba
 A^-+a^- & = & -U\p_+\delta\theta U^\dagger = 0\quad {\rm if}\,\,\,\p_+\delta\theta=0 \\
 A^i+a^i & = & \fra{i}{g} U\p_i U^\dagger + U(\tilde a^i+\p_i\delta\theta)U^\dagger \ .
\ea
Here the matrix $a_i$ is given as a product of three matrices and to compare with earlier computations we have to project out the color component $a^i_a$, using adjoint representation everywhere:
\ba
 a^i_a & = & \fra1{N_c} \Tr[T_a \, U \, T_b \, U^\dagger] (\tilde a^i+\p_i\delta\theta)_b\nn
       & = & U_{ab}(\tilde a^i_b+\p_i\delta\theta_b)=-\fra12 \epsilon_{ij}A^j_a(\epsilon,\bfx)\chi_0(0,\bfx)+U_{ab}\p_i \delta\theta_b \ .
\ea
Here the first term is what was computed in \nr{eAchi} by simply applying to the small fluctuation field the same gauge transformation $U$ as to the big background field. Transforming also $\tilde a^+$ in \nr{apsol} to zero produces the second term.

When computing the variation of $\delta\theta$ across the nuclear sheet, $0<x^-<\epsilon$, from \nr{dth} one observes that there is no singularity in the integrand so that the integral will be of the order of $\epsilon$. There is a singularity in the evaluation of $\tilde a^+$, regulated as shown in \nr{eAchi}. Taking for $A_i(y^-)$ a linear growth over the interval $0<x^-<\epsilon$ to the final value at $\epsilon$ one can estimate
\be
 \delta\theta(\epsilon)=U^\dagger(\epsilon,\bfx)A_j(\epsilon,\bfx)\fra1{\p_+}\epsilon\, [ig\,\tilde a_k(0,\bfx)\delta_{kj}+\epsilon_{kj}\,\p_k\chi_0(0,\bfx)] \ .
\ee
The contribution to the transverse gauge field induced by the gauge transformation then is $U(\epsilon,\bfx)\p_i\delta\theta$ and has two components, an axionic one inherently small and another small due to the factor $ig$. Both are small due to the overall factor $\epsilon$, reflecting the narrowness of the nuclear sheet.

\renewcommand{\theequation}{\rm{B}.\arabic{equation}}
\setcounter{equation}{0}
\section{Distribution of memory kicks}\la{aplusgauge}

We have evaluated the expectation value of the magnitude of the kick by transforming to the physical $A_i$ gauge, see Eq. \nr{p}. One must be able to do the same in the $A^+$ gauge, {\emph{i.e.}}, by integrating $p_i$ from Eq.~\nr{mshell3}. As shown by \cite{Kolbe:2020hem}, in this gauge one can also perform explicitly the averaging over color distributions and compute not only the magnitude but also their distribution. For brevity we neglect here the axionic fluctuation. 

According to \nr{derpi}, written in $A^+$ gauge in \nr{mshell3}, 
\be
 \p_- p_i(x^-)=-gQ_aF^{i+}_a(A)=-g\tilde Q_a\p_i\tilde A^+_a \la{derpi1}
\ee
with ${\bf \p}^2\tilde A^+=\tilde\rho$ solved in \nr{A+green}. In the thin sheet approximation $\tilde A^+\sim \delta(x^-)$ and one can integrate \nr{derpi1} over $x^-$ so that $\delta(x^-)$ becomes $\theta(x^-)=1$. Computing the transverse derivative of the integral representation \nr{A+green} and taking $\bfx=0$ one has the transverse kick for a fixed $\rho$:
\be
 p_i(\bfx=0)=g\tilde Q_a\int {d^2y\over2\pi}{y_i\over\bfy^2}\tilde \rho_a(\bfy)\ . \la{kickmag}
\ee 
Here $\tilde \rho$ is a transverse density of dimension two so $p_i$ has the correct dimension one. 
The distribution of kicks, normalized to 1, then is
\ba
 {dN\over d^2p} & = & {dN\over \pi dp_T^2}=\left\langle \delta^{(2)}\left(p_i-g\tilde Q_a \int {d^2x\over2\pi}{x_i\over\bfx^2}\tilde \rho_a(\bfx)\right)\right\rangle_\rho\nn
 & = & \int {d^2s\over(2\pi)^2} e^{ip_is_i}\left\langle\exp\left[-ig\tilde Q_a \int {d^2x\over2\pi}{x_is_i\over\bfx^2}\tilde \rho_a(\bfx)\right]\right\rangle_\rho \la{npi} \ .
\ea

Physics enters in the specification of the ensemble of color densities. With an exponential 
density the expectation value is
\be
 \langle \CO\rangle_\rho = \frac{\int\CD\rho_a(\bfx)  \CO(\bfx) \exp\left[-{1\over2\lambda}\sum_{a,\bfx}\, \rho_a^2(\bfx)\right]}{ \int\CD\rho_a(\bfx) \exp\left[-{ 1\over2\lambda}\sum_{a,\bfx}\, \rho_a^2(\bfx)\right]} \la{ensemble} \ .
\ee
Properties of the nuclear sheet are built in the parameter $\lambda$ of dimension 2, 
the saturation scale squared, conveniently normalized by
\be
 Q_s^2=\fra12 \lambda g^2 \tilde Q_a\tilde Q_a = \fra12 \lambda g^2  Q_a Q_a \ ,
\ee
where $Q_aQ_a\to C_A=N_c$, the adjoint Casimir. The expectation value in \nr{npi} then becomes a Gaussian integral of type 
\be
 \int \Pi dz_k\,e^{-A_k^2 z_k^2-B_kz_k}=\exp\Big[{B_k^2\over 4A_k^2}\Big]\int \Pi dz_k\,e^{-A_k^2 z_k^2} \ ,
\ee
where 
\be
 A_k^2=\fr1{2\lambda}\ ,\ B_k=ig{\tilde Q_a\over 2\pi}{s_ix_i\over\bfx^2} \ .
\ee
This yields
\be
 -{B_k^2\over 4A_k^2}=\fra12 \lambda\int d^2x {g^2\over 4\pi^2}Q_aQ_a {s_is_jx_ix_j\over \bfx^4}=\fra12 \lambda g^2Q_aQ_a{1\over 4\pi}\int{dx\over x}\,\,{\bf s}^2 ={1\over 4\pi}Q_s^2 \log{Q_s\over\Lambda}\,\,{\bf s}^2 \ .
\ee
Here we  have used the averaging $x_ix_j\to\fra12\delta_{ij}$ and regulated transverse distance integral in the IR by the QCD $\Lambda$ and in the UV by $1/Q_s$. Inserting this for the expectation value in \nr{npi} gives a Bessel function integral and
\be
 {dN\over d^2p}={1\over \pi\langle p_T^2\rangle}  \exp\left[-{p_T^2\over\langle p_T^2\rangle}\right],\quad \langle p_T^2\rangle = \langle p_i(0)p_i(0) \rangle={1\over\pi} Q_s^2 \log{Q_s\over\Lambda} \ .
\la{ptjak}
\ee
Here $\langle p_i(0)p_i(0) \rangle$ is the regulated evaluation of $\langle p_i(\bfx)p_i(\bfy) \rangle$ when $\bfy\to\bfx$, see Eq. \nr{pixpiy} (with no $x^+$ dependence). Using $p_L=p_T^2/(2M)$ from \nr{plpt} this is immediately converted to
\be
 {dN\over dp_L}={1\over \langle p_L\rangle}\exp\left[-{p_L\over \langle p_L\rangle}\right]\ , \ \langle p_L\rangle={\langle p_T^2\rangle \over 2M} \ ,
\la{pljak}
\ee
where $M$ is the mass of the test particle.

The expectation value of the magnitude of kick squared is exactly the same as the one obtained in \cite{Jokela:2019apz} by evaluating the correlator \nr{memfin}  in the $A^i$ gauge (in \cite{Jokela:2019apz} the factor $g^2$ was omitted from the correlator and $|\bfx-\bfy|$ should be regulated by $1/Q_s$). Obtaining the same result in two different gauges shows the consistency of the scheme.

\end{document}